\documentstyle[aps,twocolumn,pra,graphicx]{revtex}

\begin{document}

\draft

\wideabs{
\title{Resonator-Enhanced Optical Dipole Trap for Fermionic Lithium Atoms}

\author{A. Mosk $^a$, S. Jochim$^a$, H. Moritz$^a$,  Th. Els\"asser$^{a,c}$,
M. Weidem\"{u}ller$^a$ and R. Grimm $^b$}
\address{$^a$ Max-Planck-Institut f\"ur Kernphysik, Postfach 103980, 69029 Heidelberg, Germany
\\
{http://www.mpi-hd.mpg.de/ato/lasercool}
\\
$^b$ Institut f\"{u}r Experimentalphysik, Universit\"at
Innsbruck, 6020 Innsbruck, Austria
\\
http://exphys.uibk.ac.at/ultracold
\\
$^c$ {\footnotesize Present address: Institut f\"ur Laserphysik
Jungiusstrasse 9, 20355 Hamburg, Germany}
}
\date{May 3, 2001}
\maketitle
\begin{abstract}
We demonstrate a novel optical dipole trap which is based on
the enhancement of the optical power density of a Nd:YAG laser beam in a resonator.
The trap is particularly suited for experiments with ultracold gases, as it combines
a potential depth of order 1 mK with storage times of several tens of seconds.
We study the interactions in a gas of fermionic lithium atoms in our trap
and observe the influence of spin-changing collisions and
off-resonant photon scattering.
A key element in reaching long storage times is an ultra-low
noise laser. The dependence of the storage time on laser noise is
investigated.
\end{abstract}
OCIS: 020.7010 020.2070}

\noindent                 
Far-detuned optical dipole traps are rapidly becoming standard
tools for atomic physics at ultralow temperatures\cite{Grimm00}. They allow
trapping of practically any atomic species, and even molecules.
In the field of quantum gases
they allow trapping of mixed-state and
mixed-species ensembles.
The coupling of the atoms to the light field, which is
small for far detuned traps, can be  strongly enhanced by means of an optical resonator.
Indeed, in cavity quantum electrodynamics experiments,
 single atoms have been trapped by
a light field that corresponds to a single photon\cite{CQED}.
Optical resonators
 have
also been used to sensitively  detect
optical fields in a quantum non-demolition way
using cold atoms \cite{Sinatra98}, and even
open up new possibilities for
optical cooling of atoms and molecules\cite{Coolres}.
\begin{figure}[b]

\centering
\includegraphics[width=80mm,keepaspectratio]{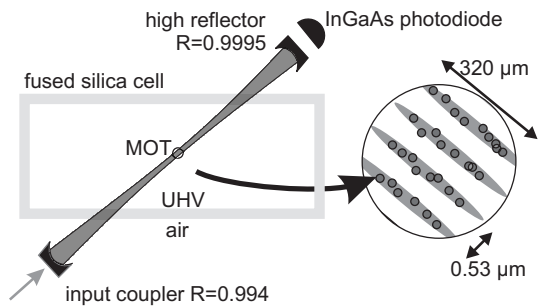}

\caption{\small Setup of the resonator trap. The atoms are trapped in
$\sim 1000$ antinodes of the standing wave.\label{figtrap}}
\end{figure}

In this letter we demonstrate a resonator-enhanced dipole trap (REDT)
for experiments with ultracold gases.
We take advantage of the resonant enhancement of the
optical power density and the corresponding trap depth. To suppress photon
scattering and reach storage times of several tens of seconds,
the detuning of the trap light from the atomic resonances is large. At the same time
the trap volume and potential depth are large to transfer many atoms into the trap.
We expect the high optical power density reached in the REDT will be useful
in many contexts, for instance
for trapping earth-alkali atoms, buffer-gas cooled atoms or
cold molecules.

Our primary interest is in spin mixtures of
fermionic $^6$Li, as a promising candidate system for the formation
of Cooper pairs in an atomic gas\cite{sympcool}. In particular,
we aim to study Feshbach
scattering resonances which have been predicted at experimentally accessible magnetic
fields, which may provide a binding mechanism for
the Cooper pairs\cite{Houbiers98}.

To sufficiently suppress photon scattering,
 the trap light must be detuned from the 670-nm D lines of Li by few
hundred nm.
In addition, to capture atoms from  a magneto-optical
trap (MOT), the optical trap must have a depth of the order of the
 temperature of Li in a MOT ($\sim 1$ mK), and a similar spatial
extension. The optical power required to create such a trap
without resonant enhancement
 exceeds
100 W.
Our REDT only requires a
 1.2 W Nd:YAG laser ($\lambda=1064 $nm), the power
 density of which
  is resonantly enhanced 130-fold
in a 15 cm near-confocal resonator
(See Fig. \ref{figtrap}). The resonator  mirrors are placed outside the vacuum, which
facilitates adjustment of the resonator and at the same
time avoids many of the problems associated with optics in ultra-high vacuum
(UHV).
 The MOT overlaps with approximately 1000 antinodes of the standing wave,
  which
act as separate microtraps, with an axial separation of $\lambda/2=532$nm
and a radial extension given by the beam waist, $w_0=160 \mu$m.
The resonator length is increased by $\sim 3$ mm from the
confocal condition to lift the degeneracy of the higher order modes.
 The optical losses at the vacuum
windows are minimized by using a high purity fused-silica UHV
cell and by traversing all intracavity glass surfaces at Brewster's
angle. The small residual round trip loss $\mathcal{L}$  permits a maximal
resonant enhancement
$
A={1}/{\mathcal{L}},
$
where the resonant enhancement factor $A$ is defined as the
increase of the intracavity intensity compared to a retroflected beam standing wave trap.
To obtain the maximum enhancement, the reflectivity $R$ of the input coupler must match $\mathcal{L}$. For our UHV cell we
  measured
 ${\mathcal{L}}=0.004(2)$ in a test resonator, and correspondingly we chose $R=0.9940(2)$.
 This
 theoretically would allow a resonant enhancement factor of 240,
 at a calculated finesse ${\cal F}=600$.
 We typically measure $A=130 \pm 15$, at ${\cal F}=650 \pm 60$, where the loss
 is due to incomplete mode matching.

 A rigid resonator body, which is acoustically decoupled from mechanical vacuum
 pumps by flexible bellows, provides passive stability.
 In addition a piezomechanical actuator
 compensates for changes in the cavity length caused e.g. by thermal drifts and acoustical
 noise. This servo loop, with a servo bandwidth of 8 kHz, uses
 the H\"ansch-Couillaud method to derive an error signal \cite{Haensch80}, where the
 Brewster windows act as the intracavity polarizer.
 No high-bandwidth stabilization proved to be necessary.
 The drive laser for the resonator trap is a commercially available ultra-low-noise
 1.2 W diode pumped solid state Nd:YAG laser (Innolight ``Mephisto'').
 Two Faraday rotators provide a $70$ dB reduction of feedback of the resonator to the
 laser, and
  an acousto-optical modulator provides control over
  the laser power admitted to the cavity. Approximately $20\%$ of the laser power is lost in
  these elements.

 The energy density inside the cavity is the same as in a retroflected 130 W beam, which
  leads to
 a  calculated trap depth of $0.8$ mK $\times k_B$ \footnote{This calculation is based on a two-level approximation
 taking into account rotating and counterrotating
 terms of the D lines of Li
 \cite{Grimm00}}.
 The corresponding photon scattering rate in the intensity maximum is calculated to be $\sim 1$
 s$^{-1}$.
In the harmonic approximation of the potential near the trap center
we calculate the axial and radial trap frequencies $\omega_{\rm ax}/2 \pi=1.4$ MHz and
 $\omega_{\rm rad}/ 2 \pi=2.0$ kHz, respectively. The trap, however,
 is very anharmonic and atoms in higher vibrational states
 oscillate at lower frequencies.

Our source of cold atoms is a MOT based on diode lasers\cite{Sch98}, which consists of
a  ``cooler'' beam, detuned $-20$ MHz from the $2^2S_{1/2}\, F=3/2 \to 2^2P_{3/2}\,
F=5/2$ transition and a ``repumper'' at $\sim -20$ MHz from
$2^2S_{1/2}\, F=1/2 \to 2^2P_{3/2}\,
F=3/2$.
In the MOT we collect approximately $2 \times  10^7$ atoms in 10 seconds from a Zeeman slowed
beam.
 The atoms then are cooled
and compressed by reducing the detunings, to a density $\sim 10^{11}$ cm$^{-3}$
 at a temperature $\lesssim
1$ mK. The MOT light beams and magnetic fields are turned
off after 20 ms of compression.
The REDT is kept permanently on during the MOT phase as it does not
influence the loading of the MOT.
Approximately $0.5\%$ of the atoms remain in the REDT after the MOT is turned off,
 the remainder being lost in the first 60 ms. Atoms remaining in
the REDT are detected by turning on the MOT fields again and
subsequently
measuring the fluorescence. The fluorescence is proportional to the
number of atoms in an optically thin MOT, with an uncertainty of $50\%$ in the
calibration.

The number of atoms transferred into the REDT can be estimated
for a shallow trap (trap depth smaller than MOT temperature)
 as the phase-space density provided by the MOT multiplied by
the number of quantum states in the shallow trap. The latter increases with the trap depth
to the power $3/2$. We observe this behavior experimentally up to
the maximum available power, even though the condition of a
shallow trap is not strictly fulfilled.

\begin{figure}[b]
\includegraphics[width=80mm,keepaspectratio]{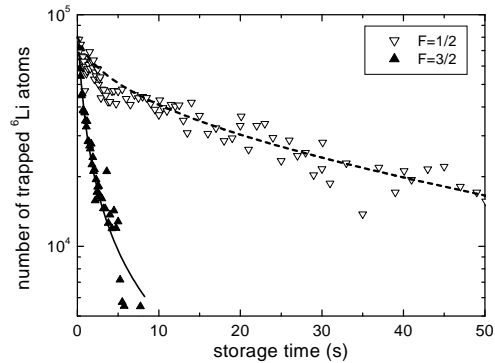}

\caption{\small Evolution of the number of trapped atoms, in
the upper (F=3/2) and lower (F=1/2) hyperfine ground states.
 Solid curve: Second order decay fit to the F=3/2 data.
Dashed curve: Modeled loss due to off-resonant photon scattering and rest gas.
\label{figtime}
}
\end{figure}
The storage time of atoms in an optical trap is usually limited by
 rest-gas collisions, interatomic processes, off-resonant photon scattering
  or heating due to laser noise.
In our apparatus, at a pressure of $ 3 \times 10^{-11}$ mbar, the decay
rate due to rest-gas collisions has been measured using
magnetically trapped Li atoms, it is $8 \times 10^{-3}$ s$^{-1}$,
which poses an upper limit to the storage time of 125 s.

Interatomic collisions in $^6$Li strongly depend on the
composition of the trapped gas.
By choosing the order in which we turn off the MOT light fields we
can influence this composition: turning off the
repumper
several milliseconds before the cooler
we obtain atoms in the $F=1/2$ state, by leaving the repumper
on longer we obtain atoms in the $F=3/2$ state. Atoms in the $F=3/2$ state
 are lost from
the trap due to spin changing collisions with a second order decay
rate $\dot{N}/N^2 = 2 \times 10^{-5}$ s$^{-1}$ (see curve in Fig. 2). Since the
 initial density of the trapped atoms
is of order  $10^{9}$ cm$^{-3}$ this implies a rate constant of
order $10^{-9}$ cm$^3/$s
 in agreement with theoretical predictions\cite{Houbiers98}.

\begin{figure}[h]
\includegraphics[width=80mm,keepaspectratio]{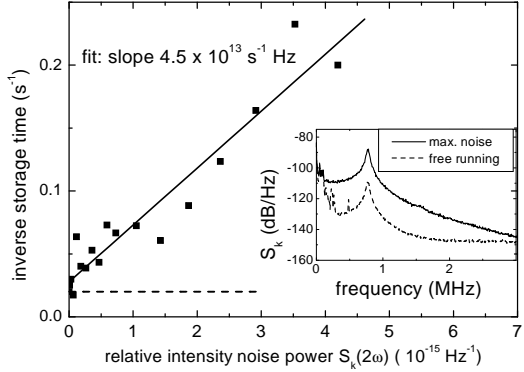}

\caption{Influence of laser intensity noise on the storage time.
 The dashed horizontal line indicates
the storage time with active stabilization of the laser output power.  The intensity noise is measured on the photodiode
 behind the resonator. Solid line: linear fit through the data.
Inset: Noise spectrum with the active intensity stabilization of the laser turned off
 (solid line)
and with additional intensity noise modulation (dashed).
Shot noise at the photodiode
exceeds the normal laser noise level.
\label{fignoise}
}
\end{figure}

The atoms in the $F=1/2$ hyperfine ground state are non-interacting
to a very good approximation, since the $s$-wave scattering length
between the two Zeeman sublevels is very small in low magnetic field
\cite{Houbiers98,OHara00}, and $p$-wave
scattering is expected to be strongly suppressed at the relevant
energies. The data show an initial fast decay on the order of 10
s, followed by a long storage time, the latter being well described by an
exponential decay with a time constant of  $\sim 50$ s. The initial decay can be
explained as follows: Since the density of trapped states in our
potential is strongly peaked just below the trap edge, a large fraction
of atoms occupy states which are only very weakly bound. A single
photon recoil momentum
then suffices to transfer such an atom into an untrapped
state, therefore the number of weakly bound atoms decays strongly
in the first few photon scattering times. We numerically modeled the effect of
 photon scattering and rest-gas collisions on the distribution of the atoms,
starting from  a  distribution  that matches the density of states in the trap.
Details of this model will be published elsewhere.
The model curve (Dashed line in Fig. 2), which has no adjustable parameters
except for the initial number of atoms, fully  agrees with our
measurements.

Laser intensity noise, especially at twice the axial trap frequency,
causes heating of the atomic gas since  fluctuations of the trap
potential at this frequency exert resonant work on the gas\cite{Bruun00}.
The resonator has a mode-cleaning effect: the modes have a
spatially well defined profile and
 pointing- and shape
fluctuations of the laser beam do not affecting the shape of the trap potential.
However, these fluctuations are
converted to intensity fluctuations.
 Great care must therefore be taken in choosing the drive laser:
  the absence of laser noise especially at frequencies around
  $2\omega_{ax}$ is crucial to obtain long storage times.
  An estimate of the relevant loss rate can be found from a
  harmonic oscillator model \cite{Gehm98}:
  $\Gamma=\pi^2 \nu^2 S_k(2\nu)$, where $\nu$ is the relevant
  oscillation
  frequency and $S_k$ is the one-sided power spectrum of relative
  intensity noise (RIN) as defined in \cite{Gehm98}.

In preliminary experiments with a different single-frequency
diode pumped Nd:YAG laser, with a
RIN of $S_k(2\omega_{\rm ax})=10^{-10}{\rm Hz}^{-1}$, we were able
to trap atoms, but storage times were less than one second.
The ``Mephisto'' laser,
 which has $S_k(2\omega_{\rm ax}) \le 10^{-14}{\rm Hz}^{-1}$,
is much more suitable as drive laser and enables the long storage
time shown in Fig. 2.
To study the strong dependence of the storage time on laser noise
we modulated the laser output power
with white noise (bandwidth 5 MHz). The resulting intracavity RIN was measured
on an InGaAs photodiode behind the resonator end mirror. Figure \ref{fignoise} shows
that
the storage time is inversely proportional to the intensity noise
level, $\tau^{-1} = (4.5 \times 10^{13} {\rm s}^{-1} )\times S_k(2\omega_{\rm ax})\times {\rm Hz}$.
The loss rate due to laser noise becomes of the same order
as the rest-gas and photon scattering terms at a RIN level of
$S_k=0.5 \times 10^{-15}{\rm Hz}^{-1}$.

The measured storage times are slightly shorter than the characteristic times
predicted by the harmonic oscillator model of Ref. \cite{Gehm98},
which is probably due to the anharmonicity of the trap: Many atoms
oscillate at lower frequencies and hence respond to a
different part of the noise spectrum.

 In conclusion, we have demonstrated a resonator-enhanced dipole
 trap that traps approximately $10^5$ fermionic lithium atoms.
 Although this type of trap is sensitive to laser noise,
  a storage time of several tens of seconds can be reached
  by using an ultra-low noise laser.
 On this time scale atoms are lost due to photon scattering and
 rest-gas collisions.

The storage time in our present system greatly exceeds the
expected collision times in the Li gas at fields $ \sim 0.08$T, which are
of order one second \cite{Houbiers98}.
 Since the thermal energy of
the trapped gas is comparable to the trap depth, these
collisions will then lead to evaporative cooling of the gas, which can be observed
 through loss of atoms. Additionally, by lowering the
trap potential in a controlled way we can measure the energy distribution
 of atoms in the trap. Thermalization and loss
measurements taken together will characterize the interactions in the vicinity of the Feshbach
resonance.

The work of A.M. is supported by a Marie-Curie fellowship from the
European Community programme IHP
under contract number CT-1999-00316. We thank  InnoLight GmbH for the loan of a
 Mephisto 1200 laser system.
 We are indebted to D. Schwalm for
encouragement and support.

\end{document}